\documentclass[11pt]{article}
\usepackage{geometry}
\title{Rank Vertex Cover as a Natural Problem for Algebraic Compression\footnote{
Supported by Parameterized Approximation, ERC Starting Grant 306992, 
and  Rigorous Theory of Preprocessing, ERC Advanced Investigator Grant 267959.}}

\author{S. M. Meesum\footnote{The Institute of Mathematical Sciences, HBNI, Chennai, India.       \texttt{\{meesum|saket\}@imsc.res.in}}
        \and Fahad Panolan\footnote{Department of Informatics, University of Bergen, Norway.         \texttt{\{fahad.panolan|meirav.zehavi\}@ii.uib.no}} \addtocounter{footnote}{-2}
       \and  Saket Saurabh\footnotemark \addtocounter{footnote}{0}~\footnotemark
\and \addtocounter{footnote}{-1} Meirav Zehavi\footnotemark
}
\date{}

\usepackage{xspace}
\usepackage{amssymb}
\usepackage{amsfonts}
\usepackage{amsthm}
\usepackage{amsmath}

\newtheorem{lemma}{Lemma}
\newtheorem{definition}{Definition}
\newtheorem{observation}{Observation}
\newtheorem{redr}{Reduction Rule}
\newtheorem{proposition}{Proposition}
\newtheorem{corollary}{Corollary}
 
\usepackage{microtype}

\usepackage{xspace}

\newcommand{\kernelsize}{$\tilde{\cO}({k^{7} +k^{4.5} \log (1/\varepsilon)})$}

\newcommand{\cO}{{\cal O}}
\newcommand{\I}{\mathcal{I}}

\newcommand{\OO}{\mathcal{O}}

\newcommand{\B}{\mathcal{B}}

\newcommand{\rank}[1]{\mbox{\sf rank}(#1)}
\newcommand{\minus}{\setminus}
\newcommand{\Yes}{{\sf Yes}}

\newcommand{\rvc}{\textsc{Rank Vertex Cover}\xspace}

\newcommand{\problemdefn}[3]{
    \begin{center}
    \noindent
    \framebox{
        \begin{minipage}{5.3in}
            \textbf{\sc #1} \\
            \emph{Input}: #2 \\
            \emph{Output}: #3
        \end{minipage}
    }
    \end{center}
}

\begin{document}

\maketitle

\begin{abstract}
The question of the existence of a polynomial kernelization of the {\sc Vertex Cover 
Above LP} problem has been a longstanding, notorious open problem in Parameterized Complexity. Five years ago, the breakthrough work by Kratsch and Wahlstr$\ddot{\mathrm{o}}$m on
representative sets has finally answered this question in the affirmative [FOCS 2012]. In this paper, we present an alternative, {\em algebraic compression} of the {\sc Vertex
Cover 
Above LP} problem into the {\sc Rank Vertex Cover} problem. Here, the input consists of a graph $G$, a parameter $k$, and a bijection between $V(G)$ and the set of columns of a
representation of a matriod $M$, and the objective is to find a vertex cover whose rank is upper bounded by $k$.
\end{abstract}

\section{Introduction}

The field of Parameterized Complexity concerns the study of {\em parameterized problems}, where each problem instance is associated with a {\em parameter} $k$ that is a non-negative integer. Given a parameterized problem of interest, which is generally computationally hard, the first, most basic question that arises asks whether the problem at hand is {\em fixed-parameter tractable (FPT)}. Here, a problem $\Pi$ is said to be FPT if it is solvable in time $f(k)\cdot|X|^{\OO(1)}$, where $f$ is an arbitrary function that depends {\em only} on $k$ and $|X|$ is the size of the input instance. In other words, the notion of FPT signifies that it is not necessary for the combinatorial explosion in the running time of an algorithm for $\Pi$ to depend on the input size, but it can be confined to the parameter $k$. Having established that a problem is FPT, the second, most basic question that follows asks whether the problem also admits a {\em polynomial kernel}.  A concept closely related to kernelization is one of {\em polynomial compression}. Here, a problem $\Pi$ is said to admit a polynomial compression if there exist a problem $\widehat{\Pi}$ and a polynomial-time algorithm such that given an instance $(X,k)$ of $\Pi$, the algorithm outputs an equivalent instance $(\widehat{X},\widehat{k})$ of $\widehat{\Pi}$,  where $|\widehat{X}|=\widehat{k}^{\OO(1)}$ and $\widehat{k}\leq k$. Roughly speaking, compression is a mathematical concept that aims to analyze preprocessing procedures in a formal, rigorous manner. We note that in case $\Pi=\widehat{\Pi}$, the problem is further said to admit a {\em polynomial kernelization}, and the output $(\widehat{X},\widehat{k})$ is called a {\em kernel}.

The {\sc Vertex Cover} problem is (arguably) the most well-studied problem in Parameterized Complexity \cite{DBLP:series/txcs/DowneyF13,ParamBook15}. Given a graph $H$ and a parameter $k$, this problem asks whether $H$ admits a vertex cover of size at most $k$.
Over the years, a notable number of algorithms have been developed for the {\sc Vertex Cover} problem~\cite{Buss93, balasubramanian1998improved, downey1999parameterized, niedermeier1999upper, Chen2001280, chandran2005refined, chen2010improved}. Currently, the best known algorithm solves this problem in the remarkable time $1.2738^k\cdot n^{\OO(1)}$~\cite{chen2010improved}. While it is not known whether the constant 1.2738 is ``close'' to optimal, it is known that unless the Exponential Time Hypothesis (ETH) fails, {\sc Vertex Cover} cannot be solved in time $2^{o(k)}\cdot n^{\OO(1)}$~\cite{impagliazzo2001problems}. On the other hand, in the context of kernelization, the picture is clear in the following sense: It is known that {\sc Vertex Cover} admits a kernel with $\OO(k^2)$ 
vertices and edges~\cite{Buss93}, but unless NP $\subseteq$ co-NP/poly, it does not admit a kernel with  $\OO(k^{2-\epsilon})$ edges~\cite{dell2014satisfiability}. We remark that it is also known that {\sc Vertex Cover} admits a kernel not only of size $\OO(k^2)$, but also with only $2k$ vertices~\cite{Chen2001280,lampis2011kernel}, and it is conjectured that this bound might be essentially tight~\cite{Chen2007}.

It has become widely accepted that {\sc Vertex Cover} is one of the most natural test beds for the development of new techniques and tools in Parameterized Complexity. Unfortunately, the vertex cover number of a graph is generally large---in fact, it is often linear in the size of the entire vertex set of the graph \cite{DBLP:series/txcs/DowneyF13,ParamBook15}. Therefore, alternative parameterizations, known as {\em above guarantee parameterizations}, have been proposed. The two most well known such parameterizations are based on the observation that the vertex cover number of a graph $H$ is at least as large as the {\em fractional vertex cover number} of $H$, which in turn is at least as large as the maximum size of a matching of $H$. Here, the fractional vertex cover number of $H$ is the solution to the linear program that minimizes $\sum_{v\in V(H)}x_v$ subject to the constraints $x_u+x_v\geq 1$ for all $\{u,v\}\in E(H)$, and $x_v\geq 0$ for all $v\in V(H)$. Accordingly, given a graph $H$ and a parameter $k$, the {\sc Vertex Cover Above MM} problem asks whether $H$ admits a vertex cover of size at most $\mu(H)+k$, where $\mu(H)$ is the maximum size of a matching of $H$, and the {\sc Vertex Cover Above LP} problem asks whether $H$ admits a vertex cover of size at most $\ell(H)+k$, where $\ell(H)$ is the fractional vertex cover number of $H$.

On the one hand, several parameterized algorithms for these two problems have been developed in the last decade~\cite{razgon2009almost,raman2011paths,CyganPPWO2013,narayanaswamy2012lp,lokshtanov2014faster}. Currently, the best known algorithm for {\sc Vertex Cover Above LP}, which is also the best known algorithm {\sc Vertex Cover Above MM}, runs in time $2.3146^k\cdot n^{\OO(1)}$ \cite{lokshtanov2014faster}. On the other hand, the question of the existence of polynomial kernelizations of these two problems has been a longstanding, notorious open problem in Parameterized Complexity. Five years ago, the breakthrough work by Kratsch and Wahlstr$\ddot{\mathrm{o}}$m on representative sets has finally answered this question in the affirmative \cite{kratsch2012representative}. Up to date, the kernelizations by Kratsch and Wahlstr$\ddot{\mathrm{o}}$m have remained the only known (randomized) polynomial compressions of {\sc Vertex Cover Above MM} and {\sc Vertex Cover Above LP}. Note that since $\ell(H)$ is necessarily at least as large as $\mu(H)$, a polynomial compression of {\sc Vertex Cover Above LP} also doubles as a polynomial compression of {\sc Vertex Cover Above MM}. We also remark that several central problems in Parameterized Complexity, such as the {\sc Odd Cycle Transversal} problem, are known to admit parameter-preserving reductions to {\sc Vertex Cover Above LP}~\cite{lokshtanov2014faster}. Hence, the significance of a polynomial compression of {\sc Vertex Cover Above LP} also stems from the observation that it simultaneously serves as a polynomial compression of additional well-known problems, and can therefore potentially establish the target problem as a natural candidate to express compressed problem instances.

Recently, a higher above-guarantee parameterization of {\sc Vertex Cover}, resulting in the {\sc Vertex Cover Above Lov\'asz-Plummer}, has been introduced by Garg and Philip \cite{GargP16}. Here, given a graph $H$ and a parameter $k$, the objective is to determine whether $H$ admits a vertex cover of size at most $(2\ell(H)-\mu(H))+k$. Garg and Philip~\cite{GargP16} showed that this problem is solvable in time $3^k\cdot n^{\OO(1)}$, and Kratsch~\cite{Kratsch16} showed that it admits a (randomized) kernelization that results in a large, yet polynomial, kernel. We remark that above-guarantee parameterizations can very easily reach bars beyond which the problem at hand is no longer FPT. For example, Gutin et al.~\cite{gutin2011vertex}
showed that the parameterization of {\sc Vertex Cover} above $m/\Delta(H)$, where $\Delta(H)$ is the maximum degree of a vertex in $H$ and $m$ is the number of edges in $H$, results in a problem that is not FPT (unless FPT=W[1]).

\medskip
\noindent
{\bf Our Results and Methods.} In this paper, we present an alternative, {\em algebraic compression} of the {\sc Vertex Cover 
Above LP} problem into the {\sc Rank Vertex Cover} problem. We remark that \rvc was originally introduced by Lov\'asz as a tool for the examination of critical
graphs~\cite{flats}. 
Given a graph $H$, a parameter $\ell$, and a bijection between $V(G)$ and the set of columns of a representation of a matroid $M$, the objective of \rvc is to find a vertex cover
of $H$  whose rank, which is defined by the set of columns corresponding to its vertices, is upper bounded by $\ell$. Note that formal definitions of the terms used in the
definition of \rvc can be found in Section~\ref{sec:prelims}.

We obtain a (randomized) polynomial compression of size \kernelsize\footnote{$\tilde{\cO}$ hide factors polynomial in $\log k$},
where $\varepsilon$ is the probability of failure. Here, by failure we mean that we output an instance of \rvc{} 
which is not equivalent to the input instance. 
In the first case, we can simply discard the output instance, and return an arbitrary instance of constant size; thus, we ensure that failure only refers to the maintenance of equivalence.
Our work makes use of properties of linear spaces and matroids, and also relies on elementary probability theory. One of the main challenges it overcomes is the conversion of the methods of Lov\'asz \cite{flats} into a procedure that works over  rationals with reasonably small binary encoding.

\section{Preliminaries}
\label{sec:prelims}

We use ${\mathbb N}$ to denote the set of natural numbers. 
For any $n\in {\mathbb N}$, we use $[n]$ as a shorthand for $\{1,2,\ldots,n\}$.
In this paper, the notation $\mathbb{F}$ will refer to a finite field of prime size.
Accordingly, $\mathbb{F}^n$ is an $n$-dimensional linear space over the field $\mathbb{F}$, where
a vector $v \in \mathbb{F}^n$ is a tuple of $n$ elements from the field $\mathbb{F}$. Here, the vector $v$ is implicitly assumed to be represented as a column vector, unless stated otherwise.
A finite set of vectors $S$ over the field $\mathbb{F}$ is said to be {\em linearly independent} if the only
solution to the equation $\sum_{v\in S} \lambda_v v = 0$, where it holds that $\lambda_v \in \mathbb{F}$ for all $v\in S$, is the one that  assigns zero to all of the scalars $\lambda_v$. A set $S$ that is not linearly independent is said to be {\em linearly dependent}.
The {\em span} of a set of vectors $S$, denoted by $\overline{S}$ (or $\rm{span}(S)$), is the set $\{ \sum_{v\in S} \alpha_v v : \alpha_v \in \mathbb{F} \}$, defined over the linear space~$\mathbb{F}^n$.

For a graph $G$, we use $V(G)$ and $E(G)$ to denote the vertex set  and the edge set of $G$, respectively.
We treat the edge set of an {\em undirected graph} $G$ as a family of subsets of size 2 of $V(G)$, i.e.~$E(G)\subseteq {V(G) \choose 2}$.
An {\em independent set} in a graph $G$ is a set of vertices $X$ such that for all $u,v \in X$, it holds that $\{u,v\} \notin E(G)$.
For a graph $G$ and a vertex $v\in V(G)$, we use $G\minus v$ to denote the graph obtained from $G$ after 
deleting $v$ and the edges incident with~$v$. 

\subsection{Matroids}
\begin{definition}
A {\em matroid} $X$ is a pair $(U,\I)$, where $U$ is a set of elements and $\I$ is a set of subsets of $U$, with the following properties:
$(i)$ $\emptyset \in \I$,
$(ii)$ if $I_1\subset I_2$ and $I_2\in \I$, then $I_1\in \I$, and
$(iii)$ if $I_1,I_2 \in \I$ and $|I_1| < |I_2|$, then there is $x \in (I_2\setminus I_1)$ such that $I_1\cup \{x\} \in \I$.
\end{definition}

A set $I' \in \I$ is said to be {\em independent}; otherwise, it is said to be {\em dependent}. A set $B\in \I$ is a {\em basis} if no superset of $B$ is independent.
For example, $U_{t,n} = ([n],\{I : I\subseteq [n], |I| \leq t \})$ forms a matroid known as a {\em uniform matroid}. 
For a matroid $X=(U,\I)$, we use $E(X),\I(X)$ and $\B(X)$ to denote the ground set $U$ of $X$, the 
set of independent sets $\I$ of $X$, and the set of bases of $X$, respectively.  
Here, 
we are mainly interested in {\em linear matroids}, which are defined as follows. Given a matroid $X=(U,\I)$, a matrix $M$ having $|U|$ columns is said to {\em represent} $X$ if {\bf (i)} the columns of $M$ are in bijection with the elements in $U$, and {\bf (ii)} a set $A\subseteq U$ is independent in $X$ if  and only if the columns corresponding to $A$ in $M$ are linearly independent. Accordingly, a matroid is a linear matroid if it has a representation over some field. For simplicity, we use the same symbol to refer to a matroid $M$ and its representation. For a matrix $M$ and some subset $B$ of columns of $M$, we let $M[\star,B]$ denote the submatrix of $M$ that is obtained from $M$ be deleting all columns not in~$B$. The submatrix of $M$ over a subset of rows $R$ and a subset of columns $B$ is denoted using $M[R,B]$. 

We proceed by stating several basic definitions related to matroids that are central to our work. For this purpose, let  $X=(U,\I)$ be a matroid.
An element $x \in U$ is called a {\em loop} if it does not belong to any independent set of $X$.
If $X$ is a linear matroid, then loops correspond to zero column vectors in its representation.
An element $x \in U$ is called a {\em co-loop} if it occurs in every basis of $X$. Note that for a linear matroid $X$, an element $x$ is a co-loop if it is linearly independent from any subset of $U\setminus \{x\}$.
For a subset $A\subseteq U$, the {\em rank} of $A$ is defined as the maximum size of an independent subset of $A$, that is,
 ${\sf rank}_X(A) := \max_{I'\subseteq A} \{|I'|: I'\in I\}$. We remove the subscript of ${\sf rank}_X(A)$, if the matroid is clear from the context.

The {\em rank function} of $X$  is the function ${\sf rank}:2^U\rightarrow \mathbb{N}$ that assigns ${\sf rank}(A)$ to each subset $A\subseteq U$. Note that this function satisfies the following properties.
\begin{enumerate}
\setlength{\itemsep}{-2pt}
 \item $0 \leq {\sf rank}(A) \leq |A|$,
\item if $A\subseteq B$, then ${\sf rank}(A) \leq {\sf rank}(B)$, and
\item ${\sf rank}(A\cup B) + {\sf rank}(A\cap B) \leq {\sf rank}(A) + {\sf rank}(B)$.
\end{enumerate}

 A subset $F\subseteq U$ is a {\em flat} if ${\sf rank}(F\cup \{x\}) > {\sf rank}(F)$ for all $x \notin F$.
Let $\cal F$ be the set of all flats of the matroid $X$. For any subset $A$ of $U$, the closure $\overline{A}$ is defined as
 $\overline{A} = \bigcap_{F\in {\cal F}} \{F: A\subseteq  F\}$.
In other words, the closure of a set is the flat of minimum rank containing it.
Let $F$ be a flat of the matroid $X$. A point $x\in F$ is said to be in {\em general position on $F$} if for any flat $F'$ of $X$, if $x$ is contained in ${\rm span}(F'\setminus\{x\})$ then $F\subseteq F'$.  


\smallskip

\noindent
{\bf Deletion and Contraction.}
The deletion of an element $u$ from $X$ results in a matroid $X'$, denoted by $X\minus u$, with ground set $E(X')=E(X)\setminus \{u\}$ 
and set of independent sets $\I(X')=\{I~:~I\in \I(X),u\notin I\}$. 
The contraction of a non-loop element $u$ from $X$ results in a matroid $X'$, denoted by $X/u$,  with ground set $E(X')=E(X)\setminus \{u\}$ 
and set of independent sets $\I(X')=\{I\setminus \{u\}~:~u\in I \mbox{ and } I \in \I(X)\}$.  Note that $B$ is a basis in $X/u$ 
if and only if $B\cup \{u\}$ is a basis in $X$. 

{\bf When we are considering two matroids $X$ and $X/u$, then for any subset $T\subseteq E(X)\setminus \{u\}$, 
$\overline{T}$ represents the closure of $T$  with respect to the matroid $X$.}

A matroid can be also be represented by a ground set and a rank function, and for our purposes, it is sometimes convenient to employ such a representation. That is, we also use a
pair $(U,r)$ to specify a matroid, where $U$ is the ground set and $r$ is rank function.  
Now, we prove several lemmata regarding operations on matroids, which are used later in the paper. 
%
%
\begin{observation}
\label{obs:contraction_reduces_rank}
Let $M$ be  a matroid, $u\in E(M)$ be  a non-loop element in $M$ and $v$ be a co-loop in $M$.  
Then, $\rank{M/u}=\rank{M}-1$ and $v$ is a co-loop in $M/u$. 
\end{observation}

Given a matrix (or a linear matroid) $A$ and a column $v\in A$, by moving the vector $v$ to another vector $u$, we refer to the operation that replaces the column $v$ by the column $u$ in $A$.

\begin{lemma}
\label{lem:isthmus-unchanged-gpos}
Let $X=(U,\I)$ be a linear matroid, $W \subseteq  U$, 
and let $u,v\notin W$ be two elements that are each a co-loop in $X$. 
Let $X'$ be the linear matroid obtained by moving $u$ to a general position on the flat spanned by $W$. 
Then, $v$ is also a co-loop in $X'$.
\end{lemma}
\begin{proof}
Let $u'$ denote the vector to which $u$ was moved (that is in general position on the span of $W$).
Notice that the only modification performed with respect to the vectors of $X$ is the update of $u$ to $u'$.
Suppose, by way of contradiction, that $v$ is not a co-loop in $X'$. Then, there exists a set of elements $S\subseteq E(X')$, where $v\notin S$, whose span contains $v$.
If $u'\notin S$, then $S\subseteq U$, which implies that $v$ was not a co-loop in $X$. Since this results in a contradiction, we have that $u'\in S$.
As $u'$ is in the span of $W$,  
$v$ must be in the span of $(W\cup S)\setminus\{u'\}$. Since $(W\cup S)\setminus \{u'\}\subseteq U$ and $v\notin (W\cup S)\setminus \{u'\}$, we have thus reached a contradiction.
\end{proof}

We remark that the proof of Lemma~\ref{lem:isthmus-unchanged-gpos} does not require the vector $u$ to be moved to a general position on the flat, but it holds true also if $u$ is moved to any vector in $\rm{span}(W)$.

\begin{lemma}
\label{lemma:isthmus-unchanged-contraction}
Let $X=(U,\I)$ be a matroid and $u\in U$ be an element that is not a loop in $X$. 
If $v\in U$ is a co-loop in $X$, 
then $v$ is also a co-loop in the contracted matroid $X/u$.
\end{lemma}
\begin{proof}
Suppose, by way of contradiction, that $v$ is not a co-loop in  $X/u$. 
Then, there exists an independent set $I \in \I(X/u)$, where $v\notin I$, whose span contains $v$. 
In particular, this implies that $I \cup \{v\}$ is a dependent set in $X/u$.
By the definition of contraction, $I\cup \{u\}$ is an independent set in $X$.
As $v$ is a co-loop in $X$, $I\cup \{u,v\}$ is also an independent set in $X$.
By the definition of contraction, $I\cup \{v\}$ is an independent set in $X/u$, which contradicts our previous conclusion 
that $I\cup \{ v\}$ is a dependent set in $X/u$.  
\end{proof}

%

\begin{lemma}[Proposition $3.9$~\cite{matroidsbook}]\label{abc}
 Let $v$ be an element in a matroid $X= (U,\I)$, which  is not a loop in $X$. Let $T$ be a subset of $U$ such that $v \in \overline{T}$. Then, ${\sf rank}_X(T) = {\sf rank}_{X/v}(T\setminus\{v\})+1$.
\end{lemma}


The lemma above can be rephrased as follows: if $T$ is a set of elements in a matroid $X=(U,\I)$ such that an element $v\in U$ is contained in the span of $T$, then the rank of $T$ in the contracted matroid  $X/v$ is smaller by $1$ than the rank of $T$ in $X$.

 \section{Compression}

Our objective is to give a polynomial compression of {\sc Vertex Cover Above LP}. More precisely, we develop a polynomial-time randomized algorithm that given an instance of {\sc Vertex Cover Above LP} with parameter $k$ and $\varepsilon>0$, with probability 
at least $1-\varepsilon$ outputs an equivalent instance of  \rvc{} whose size is bounded by a polynomial in $k$ and $\epsilon$. It is known that there is a parameter-preserving reduction from {\sc Vertex Cover Above LP} to {\sc Vertex Cover Above MM} such that the parameter of the output instance is linear in the parameter of the original instance~\cite{kratsch2012representative}. 
Thus, in order to give a polynomial compression of {\sc Vertex Cover Above LP} to \rvc\ where the size of the output instance is bounded by \kernelsize, it is enough to give a polynomial 
compression of {\sc Vertex Cover Above MM} to \rvc with the same bound on the size of the output instance. 
For a graph $H$, we use $\mu(H)$ and $\beta(H)$ to denote 
the maximum size of a matching  and the vertex cover number of $H$, respectively.
Let $(G,k)$ be an instance of {\sc Vertex Cover Above MM}.  Let $n=\vert V(G)\vert$ and $I_n$ denote the $n\times n$ identity matrix. 
That is, $I_n$ is a representation of $U_{n,n}$. Notice that $(G,k)$ is a \Yes{}-instance of   {\sc Vertex Cover Above MM} 
if and only if $(G,I_n,\mu(G)+k)$, with any arbitrary  bijection between $V(G)$ and columns of $I_n$, is a \Yes{}-instance of 
\rvc. 

In summary, to give the desired polynomial compression of {\sc Vertex Cover Above LP}, it is enough to give a 
polynomial compression of instances of the form $(G,I_n,\mu(G)+k)$ of \rvc\ where the size of the output instance is bounded by \kernelsize. Here, the 
parameter is $k$. For instances of \rvc, we assume that the columns of the 
matrix are labeled by the vertices in $V(G)$ in a manner corresponding to a bijection between the input graph and columns of the 
input matrix.  As discussed above, we again stress that now our objective is to give a polynomial compression 
of an instance of the form $(G,I_n,\mu(G)+k)$ of \rvc to \rvc, which can now roughly be thought of as a polynomial kernelization.  
We achieve the compression in two steps.  

\begin{enumerate}
\setlength{\itemsep}{-2pt}
\item In the first step, given $(G,M=I_n,\mu(G)+k)$, in polynomial time we either conclude that 
$(G,I_n,\mu(G)+k)$ is a \Yes{}-instance of \rvc\ or (with high probability of success) output an equivalent instance $(G_1,M_1,\ell)$ 
of \rvc{}  where the number of rows in $M_1$, and hence \rank{$M_1$}, is upper bounded by $\cO(k^{3/2})$. 
More over we also bound the bits required for each entry in the matrix to be $\tilde{\cO}({k^{5/2} + \log (1/\varepsilon)})$
This step is explained in Section~\ref{subsection:rankreduction}.
Notice that after this step, the graph $G_1$ 
need not be bounded by $k^{\cO(1)}$. 

\item In the second step, we work with the output $(G_1,M_1,\ell)$ of the first step, and in polynomial time we reduce the  
number of vertices and edges in the graph $G_1$ (and hence the number of columns in the matrix $M_1$). That is, output of this step is an equivalent instance  
$(G_2,M_2,\ell)$ where the size of $G_2$ is bounded by $\cO(k^{3})$. 
This step is explained in Section~\ref{subsection:graphreduction}.
%
\end{enumerate} 

Throughout the compression algorithm, we work with \rvc. Notice that the input of \rvc{} 
consists of a graph $G$, an integer $\ell$, and a linear representation $M$ of a matroid with a bijection between $V(G)$ and the set of columns of 
$M$. In the compression algorithm, we use operations that modify the graph $G$ and the matrix $M$ simultaneously. To employ  these 
``simultaneous operations''  conveniently,  we define (in Section~\ref{subsection:graphs_n_matroids}) the notion of a {\em graph-matroid pair}. We note that the definition of a graph-matroid pair is the same as a pre-geometry defined in \cite{flats},  and various lemmas from \cite{flats} which we use here are adapted to this definition. We also define deletion and contraction operations on a graph-matroid pair, and state some properties of these operations.

 \subsection{Graph-Matroid Pairs}
 \label{subsection:graphs_n_matroids}

We start with the definition of a graph-matroid pair. 
\begin{definition}
A pair $(H,M)$, where $H$ is a graph and $M$ is a matroid over the ground set $V(H)$, is called 
a {\em graph-matroid pair}. 
\end{definition}

Notice that there is natural bijection between $V(H)$ and $E(M)$, which is the identity map. 
Now,
we define deletion and contraction operations on graph-matroid pairs.

\begin{definition}
Let $P=(H,M)$ be a graph-matroid pair, and let $u\in V(H)$. The {\em deletion} of $u$ from $P$, denoted by 
$P\minus u$, results in the graph-matroid pair $(H\minus u,M\minus u)$. If $u$ is not a loop in $M$, then 
the {\em contraction} of $u$ in 
$P$, denoted by $P/u$, results in the graph-matroid pair $(H\minus u,M/u)$. For an edge $e\in E(H)$, 
$P\minus e$ represents the pair $(H\minus e,M)$
\end{definition}

We remark that matroid deletion and contraction can be done time polynomial in the size of ground set for a linear matroid. For details we refer to \cite{matroidsbook,oxley}.

\begin{definition}
Given a graph-matroid pair $P=(H, M)$, the vertex cover number of $P$ is defined as
 $\tau(P) = \min \{{\rm rank}_M(S): S \mbox{ is a vertex cover of } H\}.$ 
\end{definition}

For example, if $M$ is an identity matrix (where each element is a co-loop), then $\tau(P)$ is the vertex cover number of $H$.
Moreover, if we let $M$ be the uniform matroid $U_{t,n}$ such that $t$ is at least the size of the vertex cover number of $H$, then $\tau(P)$ again equals the vertex cover number of $H$.

Let $P=(H,M)$ be a graph-matroid pair where $M$ is a linear matroid. Recall that $M$ is also used to refer to a given linear representation of the matroid. 
For the sake of clarity, we use $v_M$ to refer explicitly to the column vector associated with a vertex $v\in V(H)$. When it is clear from context, we use $v$ and $v_M$ interchangeably.

\begin{lemma}[see Proposition $4.2$ in \cite{flats}]
\label{thm:genpos}
Let $P = (H, M)$ be a graph-matroid pair and  $v\in
 V(H)$ such that the vector $v_M$ is a co-loop in $M$, where $M$ is a linear matroid.  
Let $P'=(H,M')$ be the graph-matroid pair obtained by moving $v_M$ to a vector $v_{M'}$ in
general position on a flat containing the neighbors of $v$, $N_H(v)$. Then, $\tau(P') = \tau(P)$.
\end{lemma}

\begin{proof}
Note that the operation in the statement of the lemma does not change the graph $H$. The only change occurs in the matroid, where we map an co-loop $v_M$ to a vector lying in the span of its neighbors. It is clear that such an operation does not increase the rank of any vertex cover. Indeed, given a vertex cover $T$ of $H$, in case it excludes $v$, the rank of $T$ is the same in both $M$ and $M'$, and otherwise, since $v_M$ is an co-loop, the rank of $T$ cannot increase when $M$ is modified by replacing $v_M$ with {\em any} other vector. Thus, $\tau(P') \leq \tau(P)$.

For the other inequality, let $T$ be the set of vectors corresponding to a minimum rank vertex cover of the graph $H$ in the graph-matroid pair $P'$ (where we have replaced the vector $v_M$ by the vector $v_{M'}$). In what follows, note that as we are working with linear matroids, the closure operation is the linear span. We have the following two cases:

\paragraph*{Case 1: $v_{M'} \notin T$}
In this case 
$T$ is still a vertex cover of $H$ with the same
    rank. Thus, $\tau(P') = {\sf rank}_{M'}(T) = {\sf rank}_M (T) \geq \tau(P)$.

\paragraph*{Case 2: $v_{M'} \in T$}
Here, we have  two subcases:
   \begin{itemize}
     \item
       If $v_{M'} \notin \overline{ T \setminus{ \{v_{M'}\} } }$, then note that 
      $\tau(P')
      = {\rm rank}_{M'}(T)$ $
      = {\rm rank}_{M'}(T\setminus{\{v_{M'}\}}) + 1
      = {\rm rank}_M((T \setminus{\{v_{M'}\}}) \cup{\{v_M\}}) \geq \tau(G)$.
      The third equality follows because $v_M$ is an co-loop.
    \item
      If $v_{M'} \in \overline{T\setminus{ \{v_{M'}\} }}$, then as $v_{M'}$ is in general position on
      the flat of its neighbors, by definition this means that all of the neighbors of $v_{M'}$ are also present in
      $\overline{T\setminus{ \{v_{M'}\} }}$.
      Since $v_{M}$ and $v_{M'}$ have the same neighbors (as the graph $H$ has not been modified), all of the neighbors of $v_{M}$ belong to in $\overline{T\setminus{v_{M'}}}$.
      Thus, $\overline{T\setminus{ \{v_{M'}\} }}$ is a vertex cover of $H$.
      Therefore,
      $\tau(P') 
      = {\rm rank}_{M'}(T) = {\rm rank}_{M'}(\overline{T})
      = {\rm rank}_{M'}(\overline{T\setminus{v_{M'}}})
      = {\rm rank}_{M} (\overline{T\setminus{v_{M'}}})
      \geq \tau(P)$.
      The second equality crucially relies on the observation that rank of a set is equal to the rank of the span of the set.
\end{itemize}
This completes the proof of the lemma. 
\end{proof}

\begin{lemma}[see Proposition $4.3$ in \cite{flats}]\label{thm:contract}
 Let $P=(H, M)$ be a graph-matroid pair, and let $v$ be a vertex of $H$ that is contained in a flat spanned by its neighbors. Let $P'=P/v$. Then, $\tau(P') = \tau(P)-1$.
\end{lemma}

\begin{proof}
 Recall that the contraction of a vertex $v$ in $P$ results in the graph-matroid pair $P'
 = (H\minus v, M/v_M)$, i.e.~the vertex is deleted from the graph and contracted in the matroid. Denote the contracted matroid $M/v_M$ by $M'$.

We first prove that $\tau(P) \leq \tau(P') + 1$. Let $T$ be a minimum
    rank vertex cover in $P'$, i.e.~${\rm rank}_{M'}(T) = \tau(P')$. 
Let $W$ be a maximum sized independent set in $\I(M')$ contained in $T$. 
Then, by the definition of contraction, $W\cup \{v\}$ is a maximum sized independent set in $\I(M)$ 
contained in $T\cup \{v\}$. Moreover, $T\cup \{v\}$ is a vertex cover in $H$, and therefore we get that 
$\tau(P) \leq {\rm rank}_M(T\cup \{v\}) = \vert W\cup \{v\} \vert = {\rm rank}_M'(T)+1 = \tau(P')+1$.

Now we prove that $\tau(P') \leq \tau(P)-1$. Assume that $T$ is a minimum rank
    vertex cover of $P$. In case $v\notin T$, it holds that all of the neighbors of $v$
    must belong $T$ to cover edges incident to $v$. 
By our assumption, $v$ is in the span of its neighbors in $M$.  
Therefore, $v$ necessarily belongs to the span of $T$. 
Note that $T\setminus\{v\}$ is a vertex cover of $H'$.
    By Lemma~\ref{abc}, we have that $\tau(P) = {\rm rank}_M(T) = {\rm rank}_{M'}(T\setminus\{v\}) + 1 \geq \tau(P')+1$.
This completes the proof.
\end{proof}

 \subsection{Rank Reduction}
 \label{subsection:rankreduction}

In this section we explain the first step of our compression algorithm. Formally, 
we want to solve the following problem. 
\problemdefn{\sc Rank Reduction}{An instance $(G,M=I_n,\mu(G)+k)$ of \rvc, where $n=\vert V(G)\vert$.}{
An equivalent instance $(G',M',\ell)$ such that the number of rows in $M'$ is at most $\cO(k^{3/2})$.}
Here, we give a randomized polynomial time algorithm for {\sc Rank Reduction}. More precisely, along with the input of 
{\sc Rank Reduction}, we are given an error bound $\varepsilon>0$, and the objective is to output a ``small'' equivalent 
instance with probability at least $1-\varepsilon$.  We start with a reduction rule that reduces the rank by $2$.

\begin{redr}[Vertex Deletion]\label{rr:vd} 
Let $(P,\ell)$ be an instance of \rvc, 
where  $P=(G,M)$ is a graph-matroid pair. Let $v\in V(G)$ be a vertex such that $v_M$  is a co-loop in $M$. Let $M_1$ be the matrix obtained after moving the $v_M$ to a vector $v_{M_1}$ in general position on the flat spanned by $N_G(v)$. Let $P_1=(G,M_1)$ and let $P'=P_1/v_{M_1}$. Then output $(P',\ell-1)$ 
\end{redr}

\begin{lemma}
Reduction Rule~\ref{rr:vd} is safe. 
\end{lemma}
\begin{proof}
We need to show that  $(P,\ell)$ is a \Yes-instance if and only if $(P',\ell-1)$ is a \Yes-instance, 
 which follows from Lemmata~\ref{thm:genpos} and \ref{thm:contract}. 
\end{proof}

\begin{lemma}
\label{lem:rank_reduction}
Let $(P,\ell)$ be an instance of \rvc, 
where  $P=(G,M)$ is a graph-matroid pair. Let $(P',\ell-1)$  be the output of Reduction Rule~\ref{rr:vd}, 
where $P'=(G',M')$. Then  $\rank{M'}=\rank{M}-2$. 
\end{lemma}
\begin{proof}
In Reduction Rule~\ref{rr:vd}, we move a co-loop $v_M$ of $M$ to a vector $v_{M_1}$,
obtaining a matrix $M_1$. Note that $v_{M_1}$ lies in the span of $N_G(v)$, and therefore $v_{M_1}$ is not a co-loop in $M_1$. Hence, we have that $\rank{M_1}=\rank{M}-1$. By the definition of general position, it holds that 
$v_{M_1}$ is not a loop in $M_1$. Notice that $M'=M_1/v_{M_1}$. Therefore,  
by Observation~\ref{obs:contraction_reduces_rank}, $\rank{M'}=\rank{M_1}-1=\rank{M}-2$. 
\end{proof}

The following lemma explain how to apply Reduction Rule~\ref{rr:vd} efficiently. 
Later we will explain (Lemma~\ref{lem:prob1stepnew}) how to keep the bit length 
of each entries in the matrix bounded by polynomial in $k$.   

\begin{lemma}
\label{lem:prob1step}
Let $M$ be a linear matroid with $\vert E(M)\vert=n$ and 
let $p>2^n$ be an integer. Then, Reduction Rule~\ref{rr:vd} can be applied 
in polynomial time with success probability at least $1-\frac{2^n}{p}$.
The number of bits required for each entry in the output representation matrix is $\cO(\log p)$ times 
the number of bits required for each entry in the input representation matrix\footnote{We remark that we are unaware of a procedure to derandomize the application of Reduction Rule~\ref{rr:vd}.}.
\end{lemma}

\begin{proof}

Let $F$ be the set of columns in $M$ corresponding to $N_G(v)$. 
Using formal indeterminates $x=\{x_h: h\in F \}$, obtain a vector $g(x) =
\sum_{h \in F} x_h h$. 
Suppose the values of the indeterminates have been fixed to some numbers $x^*$ such that for any independent set $I\in M$ which does not span $F$, $I\cup \{g(x^*)\}$ is also independent. We claim that $g(x^*)$ is in general position on $F$. By definition, if $g(x^*)$ is not in general position, then there exists a flat $F'$ with $g(x^*) \in \overline{F' \setminus \{g(x^*)\}}$ but it does not contain $F$. Let $I$ be a basis of  $F'\setminus \{g(x^*)\}$, clearly $I$ does not span $F$ but $I\cup \{g(x^*)\}$ is a dependent set, which is a contradiction due to the choice of $x^*$.

Let $I$ be an independent set  which does not span $F$. 
We need to select $x$ in such a way that $D_{R,I}(x)=\det(M[R,I\cup \{g(x)\}])$ is not identically zero for some $R$. First of all, note that there is a choice of $R$ for which the polynomial $D_{R,I}(x)$ is not identically zero and has total degree one. This is so because $D_{R,I}(x) = \sum_{h\in F} x_h \det(M[R,I\cup\{h\}])$; if it is identically zero for every $R$, then $\det(M[R,I\cup \{h\}])=0$ which implies that every element $h\in F$ is spanned by $I$. Thus, this case does not arise due to the choice of $I$. If we choose $x\in [p]^{|F|}$ uniformly at random, for some number $p$, then the probability that $D_{R,I}(x) = 0$ is at most $\frac{1}{p}$ by Schwartz-Zippel Lemma. The number of independent sets in $M$, which does not span $F$, is at most $2^{n}$. By union bound, the probability that $D_{R,I}(x)=0$ for some $I$, an independent set of $M$, is at most $\frac{2^n}{p}$. Therefore, the success probability is at least $1-\frac{2^n}{p}$.

The procedure runs in polynomial time and the process of matroid contraction can at most double the matrix size, this gives us the claimed bit sizes.
\end{proof}

In the very first step of applying Reduction Rule~\ref{rr:vd}, the theorem above makes the bit sizes $\mathcal{O}(\log p)$. On applying the the rule again the bit length of entries double each time due to Gaussian elimination performed for the step of matroid contraction. This can make the numbers very large. To circumvent this, 
we show that given a linear matroid $(U,\I)$ of {\em low} rank and where the ground set $U$ is {\em small}, along with a representation matrix $M$ over 
the  field ${\mathbb R}$, for a randomly chosen {\em small} 
prime $q$, the matrix $M\mod q$ is also a linear representation of $M$ (see Lemma~\ref{lemma:modp}). 
To prove this result, we first observe that for any number $n$, the number of distinct prime factors is bounded by $\OO(\log n)$.

\begin{observation}
\label{prop:primefactor}
There is a constant $c$ such that number of distinct prime factors of any number $n$, denoted by $\omega(n)$, is at most  $c\log n$.
\end{observation}

The well-known prime number theorem implies that
\begin{proposition}
\label{prop:primecount}
There is a constant $c$ such that the number of distinct prime numbers smaller than or equal to $n$, denoted by $\pi(n)$, is  
at least $ c\frac{n}{\log n}$.
\end{proposition}


\begin{lemma}\label{lemma:modp}
Let $X=(U, \I)$ be a rank $r$ linear matroid representable by an $r\times n$ matrix $M$  
over ${\mathbb R}$ with each entry 
 between  $-n^{c'r} (1/\delta)$ and  $n^{c'r} (1/\delta)$ for some constants $c'$ and $\delta$.   
Let $\varepsilon > 0$. 
There is a number $c\in \OO(\log \frac{1}{\varepsilon})$ such that  for a prime number  $q$ chosen uniformly at random 
from the set of prime numbers smaller than or equal to   
$c \frac{n^{2r+3} (n\log n+\log(1/\delta))^2}{\varepsilon}$, the matrix $M_q = M \mod q$ over ${\mathbb R}$ represents the matroid 
$X$ with probability at least $1-\frac{\varepsilon}{n}$.
  \label{lemma:bit-reduce}
\end{lemma}
\begin{proof}
To prove that $M_q$ is a representation of $X$ (with high probability), it is enough to show that 
for any basis $B\in \B(X)$, the corresponding columns in $M_q$ are 
linearly independent. For this purpose, consider some basis $B\in \B(X)$. Since $B$ is an independent set in $M$, we have that 
the determinant of $M[\star,B]$, denoted by $\det(M[\star,B])$, is non-zero. The determinant of $M_q[\star,B]$ is equal to 
$\det(M[\star,B])\mod q$. Let $a=\det(M[\star,B])$, and let $b=a\mod q$. The value $b$ is equal to zero 
only if $q$ is prime factor of $b$. 
 Since the absolute value of each entry in $M$ is at most $n^{c'r}(1/\delta)$, 
 the absolute value of $a$ is upper bounded by $r! n^{c'r^2}(1/\delta)^r$.
By Observation~\ref{prop:primefactor}, the number of prime factors of $a$ is at most 
$c_1 (\log (r!)+c'r^2\log n+r\log(1/\delta)) $ for some constant $c_1$. The total number of bases in $X$ is at most 
$n^r$. Hence the cardinality of the set 
$F=\{z: z \mbox{ is a prime factor of } \det(M[\star,B]) \mbox{ for some }B\in \B(X) \}$
is at most
$n^r \cdot c_1 (\log (r!)+c'r^2\log n+r\log (1/\delta))
\leq
c_2 n^{r+1} (n \log n+\log(1/\delta))$ for some constant $c_2$.

By Proposition~\ref{prop:primecount}, there is a constant $c_3$ such that the number of prime factors less than or equal to   $c \frac{ n^{2r+3} (n\log n+\log (1/\delta))^2}{\varepsilon}$ is at least 
$$t=c_3 c \frac{n^{2r+3} (n\log n+\log(1/\delta))^2}{\varepsilon\log (\frac{n^{2r+3} (n\log n+\log(1/\delta))^2}{\varepsilon}) }.$$

The probability that $M_q$ is not a representation of $X$ (denote it by $M_q\not\equiv M$) is, 
\begin{eqnarray*}
\Pr[\mbox{$M_q \not\equiv M$}]&=& \Pr[q\in F] \leq \frac{\vert F\vert}{t} \\  
&\leq& \frac{c_2}{c_3 c} \cdot \frac{\log (\frac{n ^{2r+3} (n \log n+\log(1/\delta))^2}{\varepsilon}) }
{n^{r+1} (n\log n+\log(1/\delta))}\cdot  \frac{\varepsilon}{n} \end{eqnarray*}
For any $\varepsilon>0$, there is a number $c\in \OO(\log \frac{1}{\varepsilon})$ such that the above probability is at most $\frac{\varepsilon}{n}$.
This completes the proof of the lemma. 
\end{proof}

By combining Lemmata~\ref{lem:prob1step} and \ref{lemma:modp}, we can apply Reduction Rule~\ref{rr:vd}, 
such that each entry in the output representation matrix has bounded value.

\begin{lemma}
\label{lem:prob1stepnew}
Given $\varepsilon>0$, 
 Reduction Rule~\ref{rr:vd} can be applied 
in polynomial time with success probability at least $1-\frac{\varepsilon}{n}$. Moreover, each entry in the output representation matrix is at most $c \frac{n^{2r+3} (n\log n+\log(1/\epsilon))^2}{\epsilon}$, where $c\in \OO(\log \frac{1}{\epsilon})$.
\end{lemma}

\begin{proof}
Let $M$ be the input representation matrix. 
Let $\epsilon'=\varepsilon/(2n)$. Now we apply Lemma~\ref{lem:prob1step} using 
$p> \frac{2^n}{\epsilon'}$. Let $M'$ be the output representation matrix of 
Lemma~\ref{lem:prob1step}. By Lemma~\ref{lem:prob1step}, $M'$ represents $M$ 
with probability at least $1-\epsilon'$.  Observe that the absolute values of matrix entries are bounded by the value of $q$ as given in Lemma~\ref{lemma:bit-reduce}, thus
each entry in $M'$  has absolute value bounded by $n^{c' n}/\epsilon^2$ for some constant $c'$. 
Now, applying Lemma~\ref{lemma:modp} again completes the proof. 
\end{proof}
 
We would like to apply Reduction Rule~\ref{rr:vd} as may times as possible in order to obtain a ``good'' bound on the rank of the matroid. However, for this purpose, after applying Reduction Rule~\ref{rr:vd} with respect to some co-loop of the matroid, some other co-loops need to remain co-loops. Thus, 
instead of applying Reduction Rule~\ref{rr:vd} blindly, we choose vectors $v_M$ whose vertices belong to a predetermined independent set. To understand the advantage behind a more careful choice of the vectors $v_M$, suppose that we are given an independent set $U$ in the graph $G$ such that every vertex in it is a co-loop in the matroid.
Then, after we apply Reduction Rule~\ref{rr:vd} with one of the vertices in $U$, it holds that every other vertex in $U$ is still a co-loop (by Lemma~\ref{lem:isthmus-unchanged-gpos} and Observation~\ref{obs:contraction_reduces_rank}). 
In order to find a large independent set (in order to apply Reduction Rule~\ref{rr:vd} many times), we use the following two known algorithmic results about {\sc Vertex Cover Above MM}.  

\begin{lemma}[\cite{lokshtanov2014faster}]
\label{lem:algoparam}
There is a $2.3146^k\cdot n^{\OO(1)}$-time  deterministic algorithm for {\sc Vertex Cover Above MM}. 
\end{lemma}

Recall that for a graph $G$, we let $\beta(G)$ denote the vertex cover number of $G$.  

\begin{lemma}[\cite{Mishra2011}]
\label{lem:algoapprox}
For any $\epsilon>0$, there is a 
 randomized polynomial-time approximation algorithm that given a graph $G$, outputs a  
 vertex cover of $G$ of cardinality at most $\mu(G) + \cO(\sqrt{\log n})(\beta(G) - \mu(G))$, 
with  probability at least $1-\epsilon$. 
\end{lemma}

In what follows, we also need the following general lemma about linear matroids.
\begin{lemma}[\cite{matroidsbook}]\label{lemma:keeprowbasis}
 Let $M$ be an $a\times b$ matrix representing some matroid. If $M'$ is a matrix consisting of a row basis of $M$ then $M'$ represents the same matroid as $M$.
\end{lemma}

We are now ready to give the main lemma of this subsection. 
\begin{lemma}
\label{lem:deleterow}
There is a polynomial time randomized algorithm that given an instance $(G,M=I_n,\mu(G)+k)$ of \rvc{} and $\hat{\varepsilon}>0$, with probability at least $1-\hat{\varepsilon}$ outputs an equivalent instance $(G',M',\ell)$ of \rvc{}  such that the number of rows in $M'$ is at most  $\cO(k^{3/2})$. Here, $M'$ is a matrix over the field ${\mathbb R}$ where each entry is $\tilde{\cO}({k^{5/2} + \log (1/\hat{\varepsilon})})$
bits long.
\end{lemma}

\begin{proof}
Recall that $n=\vert V(G)\vert$. 
If $k\leq \log n$, then we use Lemma~\ref{lem:algoparam} to solve the problem in polynomial time. 
Next, we assume that $\log n < k$.  
Let $\delta=\hat{\varepsilon}/2$. 
Now, by using Lemma~\ref{lem:algoapprox}, in polynomial time we obtain a 
vertex cover $Y$ of $G$ of size at most $\mu(G) + c' \sqrt{\log n} \cdot k\leq \mu(G) + c' \cdot k^{3/2}$, 
with probability at least $1-\delta$, 
where $c'$ is some constant.
If we fail to compute such a vertex cover, then we output an arbitrary constant sized instance as output; and 
this will happen only with probability at most $\delta$. 
Otherwise,  
let $S=V(G)\setminus Y$. Since $Y$ is a vertex cover  of $G$, we have that
$S$ is an independent set of $G$. Hence, 
$\vert S \vert\geq n-(\mu(G) + c' \cdot k^{3/2})$. 
Since $M=I_n$, all the elements of $M$, including the ones in $S$, are co-loops in $M$. 
Now, we apply Reduction Rule~\ref{rr:vd} with the elements of $S$ (one by one). 
By Lemma~\ref{lem:isthmus-unchanged-gpos} and Observation~\ref{obs:contraction_reduces_rank}, 
after each application of Reduction Rule~\ref{rr:vd}, the remaining elements in $S$ are still co-loops. In particular, we apply Reduction Rule~\ref{rr:vd} $\vert S \vert$ many times. 
Let $(G',M',\ell)$ be the instance obtained after these $\vert S \vert$ applications of Reduction Rule~\ref{rr:vd} 
using Lemma~\ref{lem:prob1stepnew} (substituting $\varepsilon=\delta$ in  Lemma~\ref{lem:prob1stepnew}). 

By Lemma~\ref{lem:rank_reduction}, we know that after each application of Reduction Rule~\ref{rr:vd}, 
the rank reduces by $2$. Hence, 
\begin{eqnarray*}
\rank{M'}&=& \rank{M}-2\vert S \vert \\
&=&n-2\left( n-(\mu(G) + c' \cdot k^{3/2}) \right)\\
&=&-n+2\mu(G) + 2 c' \cdot k^{3/2} 
\leq   2 c' \cdot k^{3/2}  \qquad\qquad (\mbox{because }2\mu(G)\leq n).
\end{eqnarray*} 

During each application of Reduction Rule~\ref{rr:vd}, by Lemma~\ref{lemma:keeprowbasis}, we can assume that the number of rows 
in the representation matrix is exactly same as the rank of the matrix. 
Now, we return $(G',M',\ell)$ as the output. Notice that the number of rows in $M'$ 
is at most $\cO(k^{3/2})$.  

Now, we analyze the probability of success. 
As finding the approximate vertex cover $Y$ using Lemma~\ref{lem:algoapprox} 
fails with probability at most $\delta=\frac{\hat{\varepsilon}}{2}$, in order to get the required success 
probability of $1-\hat{\varepsilon}$, $\vert S \vert$ applications of  Reduction Rule~\ref{rr:vd} should 
succeed with probability at least $1-\frac{\hat{\varepsilon}}{2}$.
We suppose that the matrix $M=I_n$ is over the field ${\mathbb R}$.  
Recall that the instance $(G',M',\ell)$ is obtained after $\vert S \vert$ applications of Reduction Rule~\ref{rr:vd}.
The failure probability of each application of Reduction Rule~\ref{rr:vd} is at most $\frac{\delta}{n}$. 
Hence, by union bound the probability failure in at least one application of Reduction Rule~\ref{rr:vd} is 
at most ${\delta}$. Hence the 
total probability of success is at least $1-(\delta+\delta)=1-\hat{\varepsilon}$.  
%
%
By Lemma~\ref{lem:prob1stepnew} each entry in the output representation matrix is at most 
$c \frac{n^{2r+3} (n\log n+\log(2/\hat{\varepsilon}))^2}{\hat{\varepsilon}}$. Hence the bits required to represent an entry 
in $M'$ is at most $\tilde{\cO}({r \log n + \log (2/\hat{\varepsilon})})=\tilde{\cO}({k^{5/2} + \log (1/\hat{\varepsilon})})$. 
\end{proof}


 \subsection{Graph Reduction}
 \label{subsection:graphreduction}

In the previous subsection we have seen how to reduce the number of rows in the matroid. In this subsection 
we move to second step of our compression algorithm. That is, to reduce the size of the graph. 
Formally, we want to solve the following problem.  
\problemdefn{\sc Graph Reduction}{
An instance $(G',M,\ell)$ of \rvc{} such that the number of rows in $M$ is at most $\cO(k^{\frac{3}{2}})$}
{An equivalent instance $(G'',M',\ell)$ such that $\vert V(G'')\vert, \vert E(G'')\vert
\leq   \cO(k^3)$}
Here, we give an algorithm to reduce the number of edges in the graph. Having reduced the number of edges, we also obtain the desired bound on the number of vertices (as isolated vertices are discarded). Towards this, we first 
give some definitions and notations. In this section, we use $\mathbb{F}$ to denote either a finite field or $\mathbb{R}$.

%

\begin{definition}[Symmetric Square]
 For a set of vectors $S$ over a field $\mathbb{F}$,
 the {\em symmetric square}, denoted by $S^{(2)}$, is defined as
 $S^{(2)} = \{ u v^T + v u^T : u,v \in S \}$,
where the operation is matrix multiplication.
The elements of $S^{(2)}$ are matrices. 
We can define the rank function $r^{(2)}:S^{(2)} \rightarrow \mathbb{Z}$ by treating the matrices as ``long'' vectors over the field $\mathbb{F}$.
\end{definition}

With a rank function $r^{(2)}$, the pair $(S^{(2)}),r^{(2)})$ forms a matroid. For details we refer the reader to \cite{flats}. 

%

%
The dot product of two column vectors $a,b \in \mathbb{F}^n$ is the scalar $a^T b$ and is denoted by $\langle a, b\rangle$. 
Two properties of dot product are  $(i) \langle a, b\rangle=\langle b, a\rangle$ 
and $(ii) \langle a, b+c\rangle=\langle a, b\rangle+\langle a, c\rangle$.  

\begin{definition}
  Given a vector space $\mathbb{F}^d$ and a subspace $F$ of $\mathbb{F}^d$, the 
{\em orthogonal space} of $F$ is defined as
\(F^{\bot} = \{x\in\mathbb{F}^d~:~\langle y, x\rangle =0 \mbox{ for all } y\in F\}.\)
\end{definition}
 
The following observation can be proved using associativity of matrix multiplication 
and dot product of vectors. 
\begin{observation}
\label{obs:dotproduct}
Let $u,v,w$ be three $n$-length vectors. Then, $u v^T w= \langle v, w \rangle u$. 
\end{observation}

\begin{definition}[$2$-Tuples Meeting a Flat]
For a flat $F$ in a linear matroid $S$ (here $S$ is a set of vectors), the {\em set of $2$-tuples meeting $F$} is defined as
$F_2 := \{ u v^T + v u^T ~:~ v \in F , u \in S\}.$
\end{definition}

For the sake of completeness, we prove the following lemmata using elementary techniques from linear algebra.

\begin{lemma}[see Proposition $2.8$ in \cite{flats}]
For any flat $F$ in a linear matroid $S$ with rank function $r$, it holds that $F_2$ (the set of $2$-tuples meeting $F$) forms a flat in the matroid $(S^{(2)}, r^{(2)})$.
 \label{thm:edgedel}
\end{lemma}
\begin{proof}
Suppose, by way of contradiction, that $F_2$ is not a flat. Then, there exist $a,b \in S$ such that $e = ab^T+ba^T \in S^{(2)}$ is not in $F_2$ and
\begin{align}
 r^{(2)}(F_2 \cup \{e\}) &= r^{(2)}(F_2). \nonumber
\end{align}

As $e$ lies in the span of $F_2$, there exist scalars $\lambda_{uv}$ such that
\begin{align}\label{inspan}
 ab^T+ba^T &= \sum_{u \in F, v \in S} \lambda_{uv} (uv^T+vu^T).
\end{align}

Note that neither $a$ nor $b$ belongs to $F$, because if at least one of them belongs to $F$, then $e$ lies in $F_2$ (by the definition of $F_2$). Therefore, $F\neq S$ and it is a proper subspace of $S$, which implies that $F^\bot$ is non-empty (follows from Proposition 13.2 in~\cite{jukna2011}).
 Pick an element $x\in F^\bot$. By right multiplying the column matrix $x$ with the terms in Equation $\ref{inspan}$, we get
\begin{align}\label{prod-once}
a b^T x+b a^T x &= \sum_{u \in F, v \in S} \lambda_{uv} (u v^Tx + v u^T x) \nonumber \\
\langle b, x\rangle a+\langle a,x\rangle b &= \sum_{u \in F, v \in S} \lambda_{uv} \langle v, x\rangle u 
+ \langle u, x\rangle v\nonumber\\
&= \sum_{u \in F, v \in S} \lambda_{uv} \langle v, x\rangle u 
\end{align}

The second equality follows from Observation~\ref{obs:dotproduct}, 
and the third equality follows from the fact that $\langle u, x\rangle=0$ (because 
$u\in F$ and $x\in F^\bot$). 
Now, by taking dot product with $x$, from Equation~\ref{prod-once}, we have that
\begin{align}\label{prod-two}
 2 \langle a, x \rangle \langle b, x\rangle = \sum_{u \in F, v \in S} \lambda_{uv} \langle v, x\rangle \langle u, x \rangle = 0
\end{align}

The last equality follows from the fact that $\langle u, x\rangle=0$.
As the choice of $x$ was arbitrary, Equations~\ref{prod-once} and \ref{prod-two} hold 
for all $x\in F^\bot$.

By Equation~\ref{prod-two}, for all $x\in F^\bot$, at least one of $\langle b, x\rangle$ or $\langle a, x\rangle$ 
is zero. 
If exactly one of $\langle b, x\rangle$ or $\langle a, x\rangle$ is zero for some 
$x\in F^\bot$, then at least one of $a$ or $b$ is 
a linear combination of vectors from $F$ (by Equation~\ref{prod-once}) and hence it belongs to $F$, which is a contradiction (recall that we have argued that both $a$ and $b$ do not belong to the flat $F$). 
Now, consider the case where both $\langle b, x\rangle$ and $\langle a, x\rangle$ are zero for 
all $x\in F^\bot$.  
Then, both $a$ and $b$ belong to $F^{\bot \bot}$.
Since $F^{\bot \bot} = F$ (in the case $F$ is over a finite field, see Theorem $7.5$ in \cite{hill1986first}), again we have reached a contradiction.
\end{proof}

For a graph-matroid pair $P=(H,M)$ (here $M$ represents a set of vectors), define $\mathcal{E}(P)\subseteq M^{(2)}$ as
 $\mathcal{E}(P)=\{ u v^T + v u^T : \{u,v\} \in E(H) \}.$
Note that $\mathcal{E}(P)$ forms a matroid with the same rank function as the one of $M^{(2)}$.
Moreover, the elements of $\mathcal{E}(P)$ are in correspondence with the edges of $H$. For simplicity, we refer to an element of $\mathcal{E}(P)$ as an edge.
Using  Lemma~\ref{thm:edgedel}, 
we prove the following lemma.


\begin{lemma}[see Proposition $4.7$ in \cite{flats}]
\label{thm:symsqr}
Let $P=(H,M)$ be a graph-matroid pair, and let $r^{(2)}$ be the rank function of $\mathcal{E}(P)$. For an edge $e$ that is not a co-loop in $(\mathcal{E}(P), r^{(2)})$, it holds that $\tau(P \minus e) = \tau(P)$.
\end{lemma}
\begin{proof}
The deletion of edges cannot increase the vertex cover number, thus $\tau(P\minus e) \leq \tau(P)$. Next, we show that it also holds that $\tau(P\minus e) \geq \tau(P)$.

Let $T$ be a vertex cover of $H\minus e$. 
Notice that $\overline{T}$ is a flat in $M$. 
Denote $e=\{u,v\}$ and $F=\overline{T}$. If at least one of $u$ or $v$ lies in $F$, then 
$F$ is a vertex cover of $H$ and hence $\tau(P\minus e) \geq \tau(P)$. Hence, to conclude the proof, it is sufficient to show that at least one of $u$ or $v$ lies in $F$. Suppose, by way of contradiction, that $u,v \notin F$. Then, the edge $e=uv^T+vu^T$ does not belong to $F_2$ (the set of $2$-tuples meeting $F$). By Theorem~\ref{thm:symsqr}, we have that $F_2$ is a flat 
in $(M^{(2)},r^{(2)})$. Since $F$ is a vertex cover of $H\setminus e$, by the definition of $F_2$ and 
 $\mathcal{E}(P)$, we have that $\mathcal{E}(P) \setminus \{e\}\subseteq F_2$. 
Recall that $e$ is not a co-loop in  $(\mathcal{E}(P), r^{(2)})$. This implies that 
$e$ belongs to the closure of $\mathcal{E}(P) \setminus \{e\}$, and hence it belongs to its superset $F_2$. 
We have thus reached a contradiction. This completes the proof.  
\end{proof}

Using Lemma~\ref{thm:symsqr},  we get the following bound on the number of edges analogously to Theorem $4.6$ in \cite{flats}.

\begin{lemma}\label{thm:symsqrsize}
Let $(H,M,\ell)$ be an instance of \rvc{} and  $r=\rank{M}$. Applying the reduction given by Lemma~\ref{thm:symsqr} on $(H,M)$  exhaustively results in a graph with at most ${r+1 \choose 2}$ edges.
\end{lemma}
\begin{proof}
Observe that the dimension of the matroid $(\mathcal{E}(P),$ $r^{(2)})$ is bounded by ${r + 1 \choose 2}$, and the reduction given by Lemma \ref{thm:symsqr} deletes any edge that is not a co-loop in this matroid. In other words, once the reduction can no longer be applied, every edge is a co-loop in the matroid $(\mathcal{E}(P),r^{(2)})$, and hence the graph has at most ${r+1 \choose 2}$ edges.
\end{proof}

Lemma~\ref{thm:symsqrsize} bounds the number of edges of the graph. To bound the number of vertices 
in the graph, we apply the following simple reduction rule. 

\begin{redr}\label{rr:zvd}
Let $(G,M,\ell)$ is an instance of \rvc.  
For $v\in V(G)$ of degree 0 in $G$, output $(G\minus v,M\minus v,\ell)$.  
\end{redr}

Reduction Rule~\ref{rr:zvd} and Lemma~\ref{thm:symsqrsize} lead us to the main 
result of this subsection: 

\begin{corollary}
\label{cor:graphreduction}
There is a polynomial time algorithm, which given 
an instance $(G',M',\ell)$ of \rvc{} such that the number of rows in $M$ is at most $\cO(k^{\frac{3}{2}})$, 
outputs an equivalent instance $(G'',M'',\ell)$  such that $\vert V(G'')\vert,\vert E(G'')\vert$ $=\cO(k^3)$. Here, $M''$ is a restriction of $M'$. 
\end{corollary}

By combining both the steps  we get a polynomial compression of size 
\kernelsize{}
 for {\sc Vertex Cover Above LP}.

 \section{Conclusion}

In this paper, we presented a (randomized) polynomial compression of the {\sc Vertex Cover 
Above LP} problem into the algebraic {\sc Rank Vertex Cover} problem. With probability at least $1-\varepsilon$, the output instance is equivalent to the original instance and it is of bit length \kernelsize. Here, the probability $\varepsilon$ is part of the input. Recall that having our polynomial compression at hand, one also obtains polynomial compressions of additional well-known problems, such as the {\sc Odd Cycle Transversal} problem, into the {\sc Rank Vertex Cover} problem.

Finally, we note that we do not know how to derandomize our polynomial compression, and it is also not known how to derandomize the polynomial kernelization by Kratsch and Wahlstr$\ddot{\mathrm{o}}$m \cite{kratsch2012representative}.
Thus, to conclude our paper, we would like to pose the following intriguing open problem: Does there exist a deterministic polynomial compression of the {\sc Vertex Cover 
Above LP} problem?

%



\bibliography{ref}

\end{document}